\pdfoutput=1
\documentclass[11pt,twoside,a4paper,cmspaper,final,collab]{cms-tdr}
\def\svnVersion{27f7bcb}\def\svnDate{2023/09/01}\def\cmsCernNoTag{CERN-EP-2023-071}\def\cmsCernDate{\today}\def\cmsMessage{Published in Physical Review Letters as \href{http://dx.doi.org/10.1103/PhysRevLett.131.091903}{\doi{10.1103/PhysRevLett.131.091903}.}}
\begin{document}\cmsNoteHeader{BPH-22-003}

\newcommand{\e}{\ensuremath{\PGh}\xspace}
\newcommand{\ep}{\ensuremath{\PGhpr}\xspace}
\newcommand{\emmmm}{\ensuremath{\e \to \PGmp \PGmm \PGmp \PGmm}\xspace}
\newcommand{\eppmm}{\ensuremath{\e \to \PGpp \PGpm \PGmp \PGmm}\xspace}
\newcommand{\emm}{\ensuremath{\e \to \PGmp \PGmm}\xspace}
\newcommand{\four}{\ensuremath{\e \to 4 \PGm}\xspace}
\newcommand{\two}{\ensuremath{\e \to 2 \PGm}\xspace}
\newcommand{\Ntwomu}{\ensuremath{N_{2 \PGm}^{\mathrm{i,j}}}\xspace}
\newcommand{\Nfourmu}{\ensuremath{N_{4 \PGm}}\xspace}
\newcommand{\Btwomu}{\ensuremath{\mathcal{B}_{2 \PGm}}\xspace}
\newcommand{\Bfourmu}{\ensuremath{\mathcal{B}_{4 \PGm}}\xspace}
\newcommand{\experr}{\ensuremath{\,\text{(}\Btwomu\text{)}}\xspace}
\newcommand{\pttwomu}{\ensuremath{\pt^{2 \PGm}}\xspace}
\newcommand{\ptfourmu}{\ensuremath{\pt^{4 \PGm}}\xspace}
\newcommand{\mtwomu}{\ensuremath{m_{2 \PGm}}\xspace}
\newcommand{\mfourmu}{\ensuremath{m_{4 \PGm}}\xspace}
\newcommand{\Atwomu}{\ensuremath{A_{2 \PGm}^{\mathrm{i,j}}}\xspace}
\newcommand{\Afourmu}{\ensuremath{A_{4 \PGm}^{\mathrm{i,j}}}\xspace}
\ifthenelse{\boolean{cms@external}}{\providecommand{\prlWidth}{0.95\columnwidth}}{\providecommand{\prlWidth}{0.6\textwidth}}

\cmsNoteHeader{BPH-22-003}

\title{Observation of the rare decay of the \texorpdfstring{\e}{eta} meson to four muons}

\date{\today}

\abstract{
A search for the rare \emmmm double-Dalitz decay is performed using a sample of proton-proton collisions, collected by the CMS experiment at the CERN LHC with high-rate muon triggers in 2017--2018 and corresponding to an integrated luminosity of 101\fbinv. A signal having a statistical significance well in excess of 5 standard deviations is observed. Using the \emm decay as normalization, the branching fraction $\mathcal{B}(\emmmm) = [ 5.0 \pm 0.8 \stat \pm 0.7 \syst \pm 0.7 \experr ] \times 10^{-9}$ is measured, where the last term is the uncertainty in the normalization channel branching fraction. This work achieves an improved precision of over five orders of magnitude compared to previous results, leading to the first measurement of this branching fraction, which is found to agree with theoretical predictions.
}

\hypersetup{
pdfauthor={CMS Collaboration},
pdftitle={Observation of the rare decay of the eta meson to four muons},
pdfsubject={CMS},
pdfkeywords={CMS, eta meson, double Dalitz decay, scouting, rare decays, high-rate triggers}}

\maketitle 

\begin{figure}[htb]
\centering
\includegraphics[width=\prlWidth]{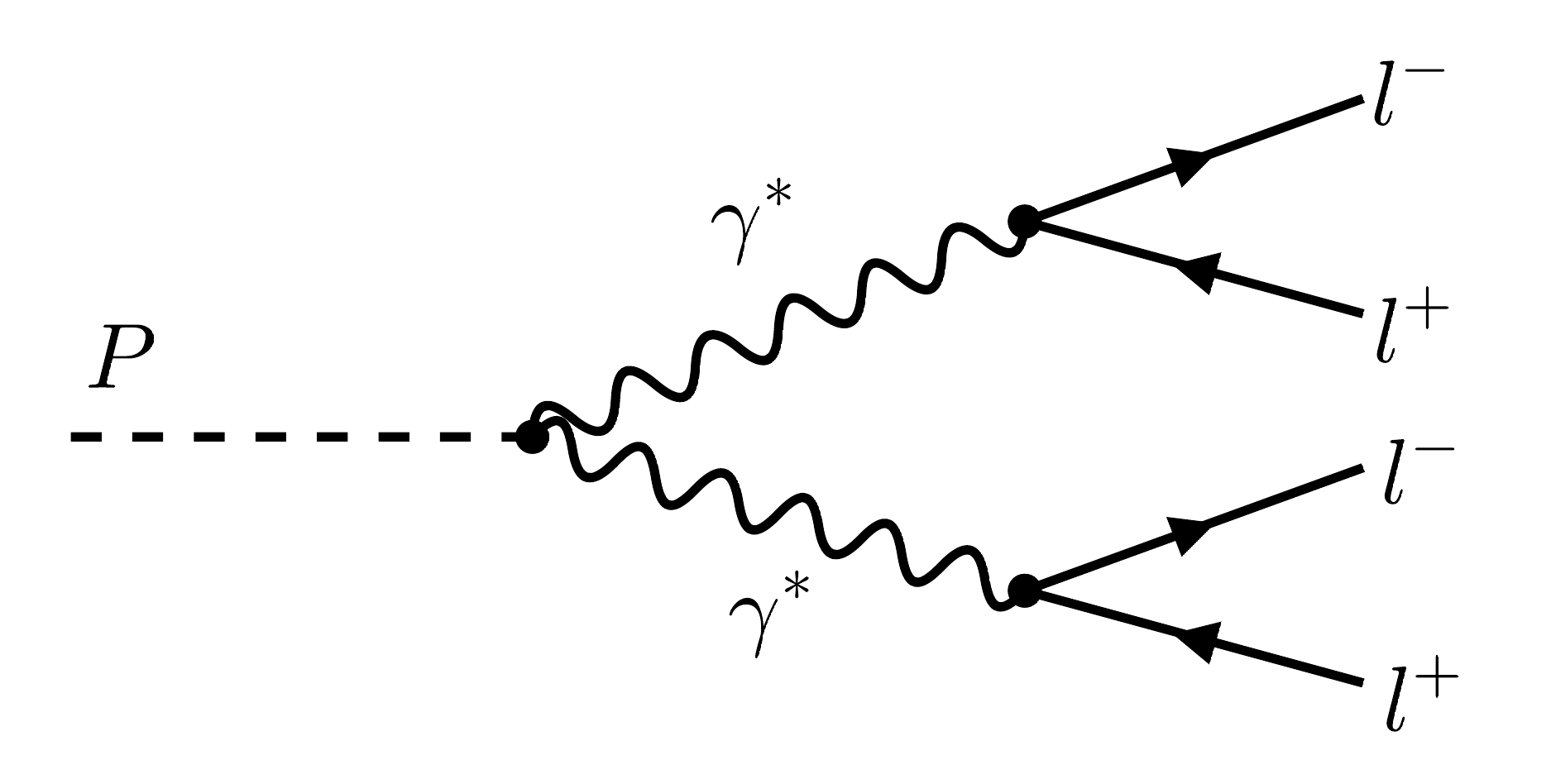}
\caption{Feynman diagram of pseudoscalar decays into four leptons, known as double-Dalitz decays.}
\label{fig:feynman}
\end{figure}

The \e and \ep mesons are $J^{PC} = 0^{-+}$ particles with masses of 547.9 and 957.8\MeV, respectively, comprising admixtures of up, down, 
and strange quarks~\cite{PDG2022}. Despite a comprehensive experimental
campaign~\cite{LANDSBERG1985301,abeggMeasurementBranchingRatio1994,ambrosinoObservationRareEta2011,etaprimeto4e,etato3pi} to study these light
mesons, several properties of the \e and \ep remain unmeasured. Their leptonic radiative decays, also known as Dalitz decays, constitute such an
example. They proceed via the electromagnetic coupling of pseudoscalar mesons to the photon, where one or more of the photons internally convert 
into a pair of leptons, as shown in Fig.~\ref{fig:feynman}. Such decays are typically highly suppressed because they can only occur through these 
electromagnetic interactions instead of the comparatively stronger nuclear interactions. To date, the only observed leptonic radiative decays are 
\emm, $\e \to \Pep \Pem \Pep \Pem$, and more recently 
$\ep \to \Pep \Pem \Pep \Pem $~\cite{abeggMeasurementBranchingRatio1994,ambrosinoObservationRareEta2011,etaprimeto4e}. 
The decays $\e \to \Pep \Pem$, \emmmm, $\e \to \Pep \Pem \PGmp \PGmm$, and most \ep decays have so far eluded discovery. 
Observing these rare decays is important because they can serve as precision tests of the standard model, they offer sensitivity to an array of new 
physics scenarios~\cite{Gan:2020aco,goudzovskiNewPhysicsSearches2022}, and the interaction between pseudoscalars and photons contributes to the 
hadronic light-by-light component of the anomalous magnetic moment of the muon~\cite{knechtScalarMesonContributions2018,Gan:2020aco}. 
A thorough description of radiative decays and their impact on several relevant standard model observables, as well as their sensitivity 
to new physics, can be found in Ref.~\cite{goudzovskiNewPhysicsSearches2022}.

In this Letter, we report the first observation of the muon double-Dalitz decay of the \e meson, \emmmm, 
using $\sqrt{s} = 13\TeV$ proton-proton ($\Pp\Pp$) collision data collected by the CMS experiment during 2017 and 2018 at the LHC. 
The predicted branching fraction for this decay channel is extremely small, 
$\mathcal{B}(\four) = (3.98 \pm 0.15) \times 10^{-9}$~\cite{escribanoDatadrivenApproachPi2018}, making its observation particularly 
challenging. The analysis measures the rate of \four events compared to the normalization channel \two~\cite{abeggMeasurementBranchingRatio1994}, 
for which the branching fraction is known with a precision of 14\%~\cite{PDG2022}. The product of the CMS detector acceptance and signal 
efficiency, as well as systematic uncertainties, are evaluated with simulation studies for both channels. Tabulated results are 
provided in the HEPData record for this analysis~\cite{hepdata}. The defining feature of the measurement is the use of high-rate triggers, 
which extends the sensitivity of CMS to dimuon and four-muon resonances of masses lower than what is achievable with the 
standard muon triggers~\cite{CMS:2016ngn}, leading to an improved precision of five orders of magnitude compared to previous works~\cite{PhysRevD.77.032004}.

The CMS apparatus is a multipurpose detector designed to trigger on and identify electrons, muons, photons, and (charged and neutral) 
hadrons~\cite{CMS:2020uim,CMS:2018rym,CMS:2014pgm}. A superconducting solenoid of 6\unit{m} internal diameter provides a magnetic 
field of 3.8\unit{T}. Within the solenoid volume are the silicon pixel and strip tracker, a crystal electromagnetic calorimeter, 
and a brass and scintillator hadron calorimeter. Muons are measured in gas-ionization detectors embedded in the steel flux-return 
yoke outside the solenoid, covering the pseudorapidity range $\abs{\eta} < 2.4$. A more detailed description of the CMS detector, 
together with a definition of the coordinate system used and the relevant kinematic variables, can be found in Ref.~\cite{CMS:2008xjf}.

Events are first selected by the two-tiered trigger system~\cite{CMS:2016ngn}. The first tier (level-1, or L1) is a hardware-based 
trigger and uses information from the calorimeters and the muon detectors to select events at a rate of roughly 100\unit{kHz}. 
The second tier (high-level trigger, or HLT) consists of a farm of processors running the CMS reconstruction software optimized 
for fast online processing. The HLT processes events at an output rate of ${\sim}1\unit{kHz}$ with approximately 1\unit{MB/event}, 
a limit imposed by the total available data transfer bandwidth. These are the ``standard triggers''~\cite{sirunyanSearchNarrowResonance2020}. 
The standard dimuon triggers require one muon with transverse momentum \pt larger than $12 \, (15)\GeV$ at L1 in $2017 \, (2018)$ and 
17\GeV at HLT, and a second muon with \pt at least $5 \, (7)\GeV$ at L1 in $2017 \, (2018)$ and 8\GeV at HLT. These triggers collect 
data at a rate of about 30\unit{Hz} at a peak instantaneous luminosity of $2\times10^{34}\unit{\percms}$~\cite{sirunyanSearchNarrowResonance2020}.

For dimuon resonance masses below ${\sim}40\GeV$, the performance of the standard muon triggers deteriorates. To improve efficiency 
for those masses, a dedicated set of high-rate dimuon triggers was developed. These triggers allow considerably lower muon \pt thresholds, 
but store only a limited amount of information per event, ensuring that the total data transfer bandwidth remains affordable. 
The information stored in the event includes only muons reconstructed at the HLT, apart from additional limited event-level information, 
leading to an event size of around 4\unit{kB} in 2017 and 8\unit{kB} in 2018~\cite{sirunyanSearchNarrowResonance2020}. This strategy, 
first introduced by the CMS Collaboration in 2012, is referred to as ``data scouting''~\cite{CMS-DP-2012-022,PhysRevLett.117.031802,sirunyanSearchNarrowResonance2020}.

The high-rate triggers include several requirements at L1, described in Table~\ref{tab:L1}. The fraction of events passing the 
full four-muon selection that is accepted by each path is shown on the last column. The first L1 path (\#1) collects most of 
the events of interest for the dimuon invariant mass (\mtwomu) region below about 1\GeV; the decay products of the \e meson 
are generally collimated, given the small mass of the \e meson and its comparatively high \pt distribution. At the HLT, each 
of the two muons is further required to have a \pt of at least 3\GeV. These high-rate triggers were commissioned in 2017 and 
found to provide a rate of ${\sim}2\unit{kHz}$ at the peak instantaneous luminosity of 
$2\times10^{34}\unit{\percms}$~\cite{sirunyanSearchNarrowResonance2020}, more than 60 times the rate of the 
standard muon triggers. The data were collected in 2017 and 2018, corresponding to an integrated luminosity of 
101\fbinv~\cite{CMS:LUM-17-003,CMS:LUM-17-004,CMS:LUM-18-002}.

\begin{table}[htb]
\topcaption{Set of dimuon L1 requirements applied in the high-rate triggers. The angular separation between muons in the 
$\eta$-$\phi$ plane is defined as $\Delta R \equiv \sqrt{\smash[b]{(\Delta\eta)^2+(\Delta\phi)^2}}$, where $\phi$ is the 
azimuthal angle. The first L1 path requires for each muon $\pt > 4.0 \, (4.5)\GeV$ in 2017 (2018). The third path imposes 
separate requirements on each muon \pt. The \mtwomu column denotes a range of allowed dimuon invariant masses. The charge 
column indicates if an opposite-sign (OS) requirement exists between muons. The last column reports the fraction of selected events accepted by each path.}
\def\arraystretch{1.3}
\label{tab:L1}
\centering\begin{scotch}{cccccccc}
L1 path & \pt [\GeVns{}]        & $\abs{\eta}$  & $\Delta R$    & \mtwomu [\GeVns{}]    & Charge    & Fraction \\
\hline
\#1     & $>$4.0\,(4.5) & \NA            & $<$1.2        & \NA                & OS        & 90\% \\
\#2     & \NA                & $<1.5$        & $<1.4$        & \NA                & OS        & 48\% \\
\#3     & $>$15,\,${>}$7        & \NA            & \NA            & \NA                & \NA        & 46\% \\
\#4     & $>$4.5            & $<2.0$        & \NA            & 7--18             & OS        & 9\% \\
\end{scotch}
\end{table}

Events passing the high-rate trigger requirements are further selected to improve the signal-to-background discrimination. 
The \two selection requires two oppositely charged muons consistent with production at a common origin (\ie, vertex). 
The vertex reconstruction is based on a Kalman filtering algorithm~\cite{fruhwirthApplicationKalmanFiltering1987a} and done 
pairwise, regardless of the decay channel. The \four selection requires four muons with a net charge of zero that are 
also consistent with same-vertex production. In this case, it means that the full set of six dimuon vertices must be reconstructed 
and compatible in position. In both two- and four-muon selections, more than one combination of muons (``candidates'') per event is 
allowed if all muons satisfy the listed requirements, but in both cases more than 98\% of events in the signal mass window 
0.53--0.57\GeV feature only one passing combination. To maximize the signal efficiency for reconstructing and identifying 
candidate \e meson decays, no additional selection is applied.

\begin{figure}[hbt]
\centering
\includegraphics[width=\prlWidth]{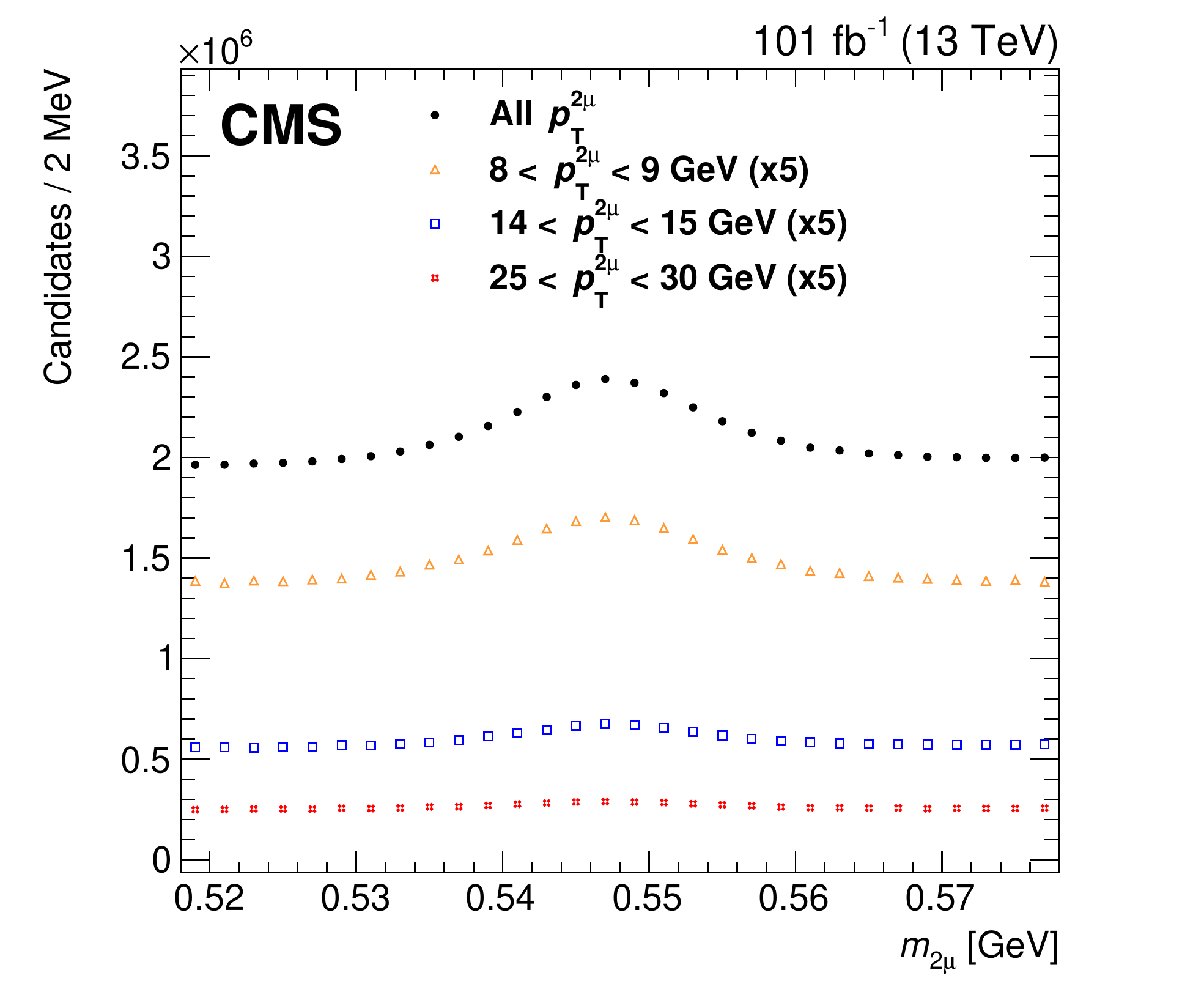}
\caption{Distribution of \mtwomu obtained with the dimuon selection integrated in dimuon \pt and in three \pt 
ranges as indicated in the legend, with the number of events in the selected \pt ranges multiplied by five for better visibility.}
\label{fig:twoMuData}
\end{figure}

Figure~\ref{fig:twoMuData} shows the \mtwomu spectrum in the region around the \e meson mass, 0.518--0.578\GeV, obtained with 
data collected using the high-rate triggers after applying the two-muon selection. The distributions are shown for the inclusive 
dimuon \pt (\pttwomu) range and additionally for three slices, 8--9, 14--15, and 25--30\GeV, highlighting the rapid decrease of 
the production cross section with \pttwomu. A binned maximum likelihood fit to the total spectrum with a sum of two Gaussian functions 
and a third-order Chebychev polynomial reveals approximately 4.5 million \two decays. Given the branching fraction 
$\mathcal{B}(\two) = (5.8 \pm 0.8) \times 10^{-6}$~\cite{PDG2022}, the total number of \e mesons produced in the sensitive kinematic 
region of the CMS detector is ${\sim}10^{12}$, neglecting reconstruction inefficiencies. This huge \e production rate is 
critical to the study of its rare decay channels.

\begin{figure}[htb]
\centering
\includegraphics[width=\prlWidth]{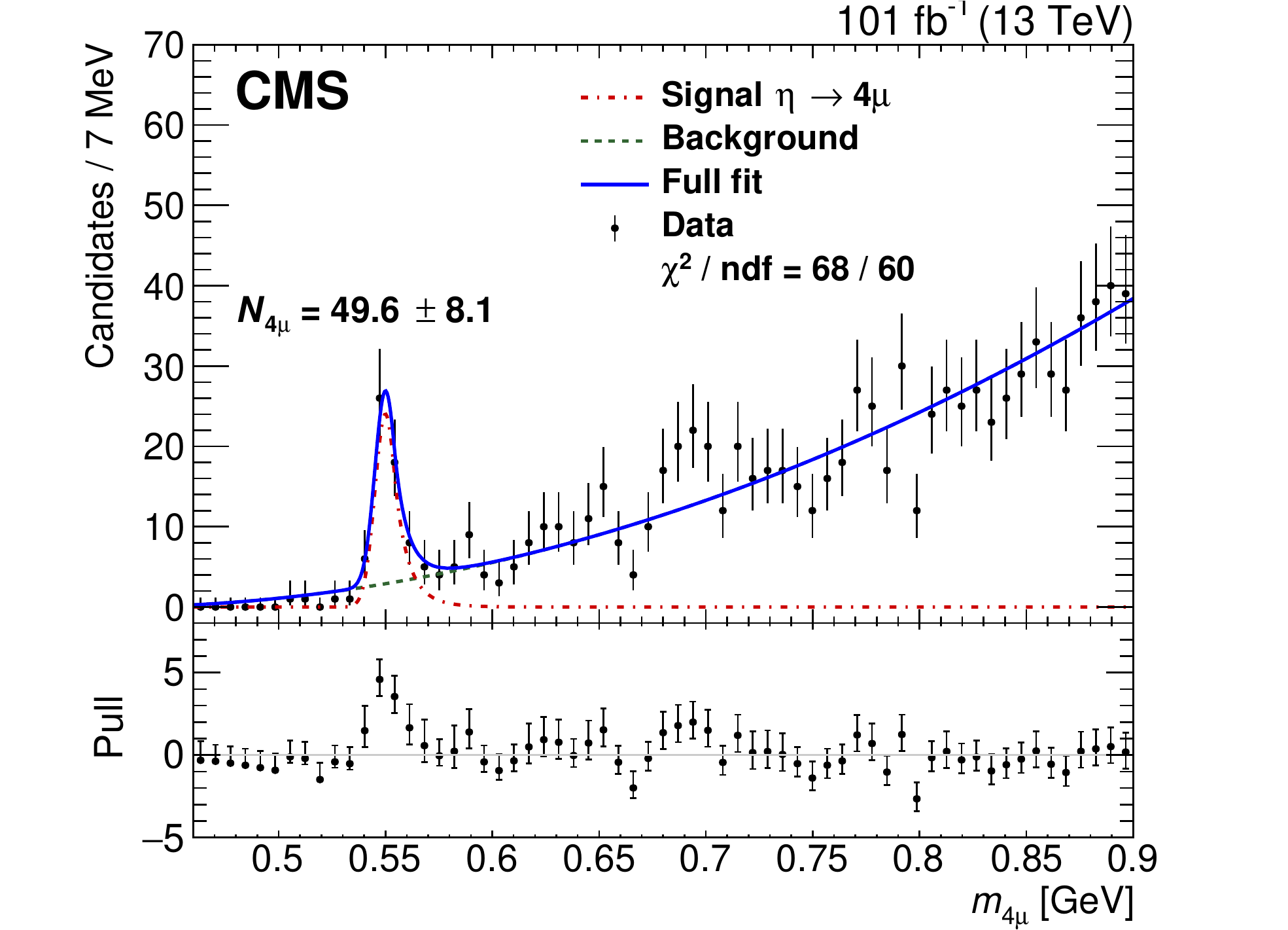}
\caption{Measured \mfourmu distribution, with the fit result overlaid. The pull distribution in the lower panel is shown relative 
to the background component of the fit model and defined as 
$(\text{Data} - \text{FitBkg})/\sqrt{\sigma_{\text{Data}}^2 - \sigma_{\text{FitBkg}}^2}$. Uncertainties are statistical only.}
\label{fig:fourMuonMassFitData}
\end{figure}

Figure~\ref{fig:fourMuonMassFitData} shows the four-muon invariant mass (\mfourmu) distribution in the range 0.46--0.90\GeV, 
obtained with data collected using the high-rate triggers after applying the four-muon selection. A clear peak is observed in the 
signal mass window 0.53--0.57\GeV, corresponding to the \e meson mass. A binned maximum likelihood fit of the spectrum to the sum 
of a single-sided Crystal-Ball function~\cite{bib:CB} for the signal and a threshold function proportional to $(\mfourmu - 4m_{\PGm})^\beta$ 
for the background, where $m_{\PGm}$ is the muon mass and $\beta$ is a free parameter of the fit, yields $\Nfourmu = 49.6 \pm 8.1$ signal 
events and $16.6 \pm 0.6$ background events (in the signal mass window). The parameters of the signal function, except for the normalization, 
are fixed from simulation. This corresponds to a statistical significance in excess of 5 standard deviations, determined by means of a 
log likelihood ratio test with a saturated model~\cite{BAKER1984437}, under a signal-plus-background (S+B) hypothesis compared 
to the background-only assumption.

Simulated samples of rare \e decays are generated at leading order with a custom workflow. The first step employs the $\textsc{Pluto v6}$ 
generator~\cite{Frohlich:2007bi} to simulate the two- and four-muon decays of the \e meson in its rest frame, using the vector 
meson dominance model~\cite{SAKURAI19601}. Subsequently, the \e meson and its decay products are boosted to the laboratory frame 
by sampling from uniform \pt and rapidity distributions in the 5--70\GeV and $\abs{y} < 2.4$ ranges, respectively. The decay 
products are then embedded into complete CMS events, which also include the simulation of fragmentation, parton shower, and 
hadronization processes in the initial and final states with the \PYTHIA\,8.230 package~\cite{Sjostrand:2014zea}, and simulation 
of the underlying event with the CP5 tune~\cite{thecmscollaborationExtractionValidationNew2020}. The location of the decay products 
in the detector is sampled from the distribution of the beam envelope. Finally, the interaction of final-state particles with the 
CMS detector is simulated using the \GEANTfour toolkit~\cite{Agostinelli:2002hh}. Simulated events include the contribution of additional 
particles produced in time within the same or nearby bunch crossings. The multiplicity of vertices is matched to the one observed in the data.

\begin{figure}[hbt]
\centering
\includegraphics[width=\prlWidth]{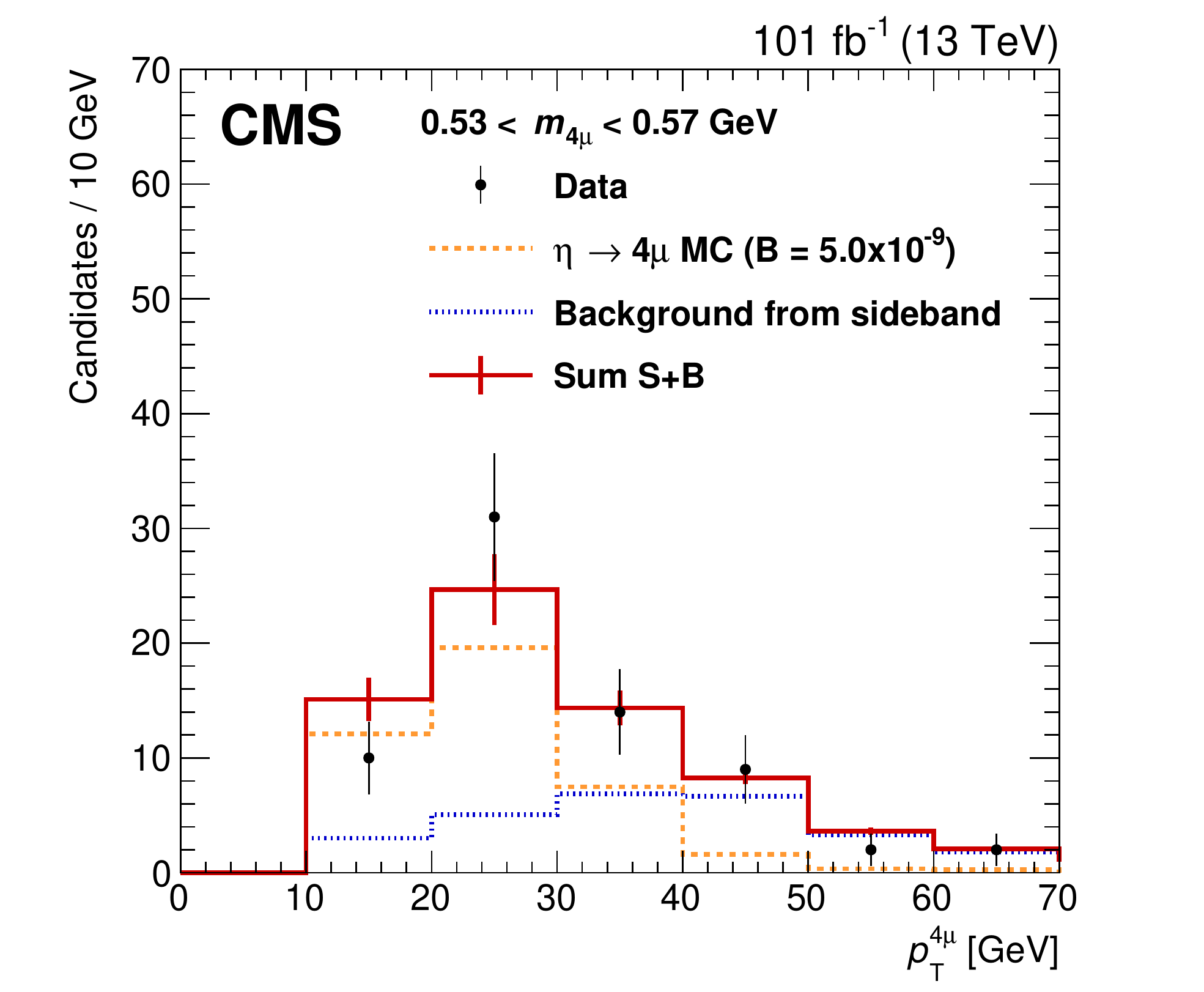}
\caption{Comparison of the four-muon \pt spectrum observed in data with $0.53 < \mfourmu < 0.57 \GeV$ (black points) and the signal 
prediction from simulation assuming the observed branching fraction (orange dashed line). Also shown is the predicted background shape, 
extracted from the mass sideband, normalized to the background fit yield (dotted blue line) and the sum of signal and background predictions (solid red line).} 
\label{fig:ExpVsObsYields}
\end{figure}

To check whether the observed peak is indeed compatible with the \four decay, the signal simulation was used to predict the four-muon 
\pt (\ptfourmu) spectrum for the experimentally measured branching fraction $\mathcal{B}(\four)$ of $5 \times 10^{-9}$ (described later). 
The signal \ptfourmu distribution was reweighted based on the \e meson \pt differential production rate measured with the two-muon channel. 
Figure~\ref{fig:ExpVsObsYields} compares the predicted distribution to the measured spectrum and to the expected background, 
obtained from events with \mfourmu within the 0.6--0.9\GeV sideband, where no signal is expected. The shape is then normalized such 
that the total background yield is fixed to the one extracted from the \mfourmu fit. The correlation between \ptfourmu and \mfourmu 
for background events was verified to be weak by comparing the \mfourmu spectrum across several \ptfourmu ranges. The sum of the 
predicted signal and background contributions agrees with the observation. An additional check was performed by applying a tight 
muon selection to muons in the signal mass window and in the sideband, and comparing the fraction of muons surviving the selection 
between the two regions. This selection requires at least six energy deposits per muon track in the silicon tracker, of which at least 
one is in the pixel layers, and a track $\chi^2/\mathrm{ndf} < 10$, where $\mathrm{ndf}$ is the number of degrees of freedom. 
About 99\% of four-muon combinations passed this selection in the signal mass window, compared to only 84\% in the sideband. 
Since the sidebands presumably contain more hadrons misidentified as muons, this indicates a negligible contamination of such hadrons in the signal region.

\begin{figure}[hbt]
\centering
\includegraphics[width=\prlWidth]{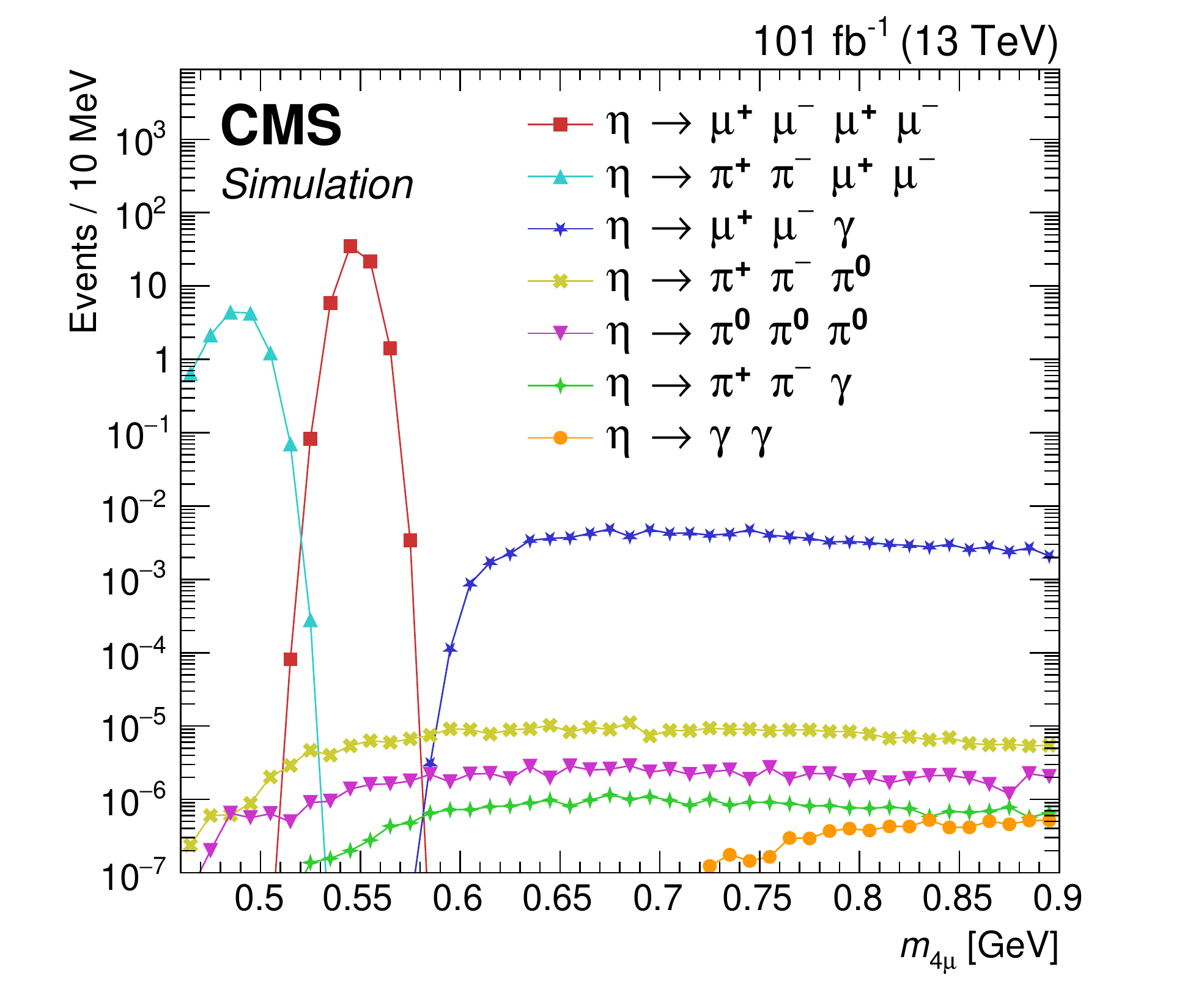}
\caption{Predicted background contributions to the signal mass window, estimated with simplified MC simulations. 
The \emmmm signal is displayed as a benchmark (red squares), followed by various other decay modes of 
the \e meson. The curves are normalized to an integrated luminosity of 101\fbinv. For unobserved decay modes, the current experimental 
upper limits are conservatively taken as the reference branching fractions in the estimations, with the exception of the signal channel, 
where the branching fraction measured in this work is used.} 
\label{fig:resonant_bkgs}
\end{figure}

To ensure that the observed peak cannot be explained by other resonant backgrounds, such as additional decay modes of the \e meson, 
we studied these modes using simplified Monte Carlo (MC) simulations. The decay chains are simulated via a series of two-body 
decays of the form $M \to m_1 m_2$. The results are reported in Fig.~\ref{fig:resonant_bkgs}. Decays with final-state photons 
(or neutral pions) could mimic the four-muon signal if a photon converts to a pair of muons after interacting with detector material. 
However, in this case, the nearby nucleus imparts a momentum kick to the system and increases its total invariant mass. Therefore, as 
a general feature of this topology, backgrounds involving photons do not present a peak in the signal mass window. The only peaking 
background is \eppmm, where both pions are misidentified as muons. Here, however, the difference in mass 
between the two particles shifts the invariant mass peak down to about 0.48\GeV, sufficiently below the signal mass window. Furthermore, 
this decay mode has never been observed. The current experimental upper limit of $\mathcal{B}(\eppmm) < 1.6 \times 10^{-4}$ 
at 90\% confidence level~\cite{PhysRevD.77.032004}, which is considerably higher than the theoretical prediction of 
$6.5 \times 10^{-9}$~\cite{borasoyEtaEtaPrime2007}, is used in our estimate. Based on these studies, we conclude that other decay modes 
of the \e meson provide a negligible contribution to the selected events.

\begin{figure}[h!]
\centering
\includegraphics[width=\prlWidth]{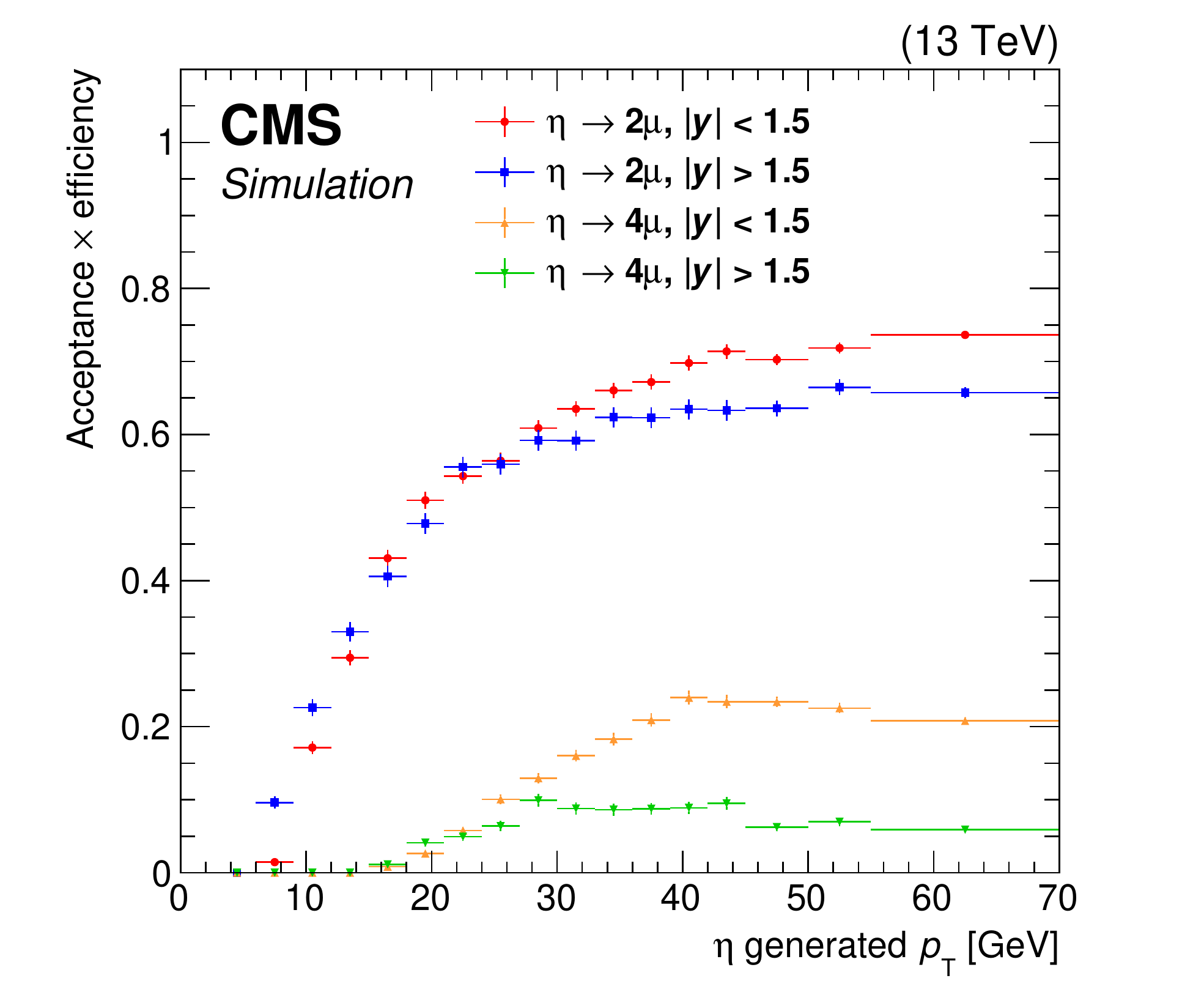}
\caption{The total efficiencies for the four-muon (\Afourmu, red and blue points) and two-muon (\Atwomu, orange and green points) 
decay channels, as functions of the generated meson's \pt and $y$, evaluated through MC simulation.}
\label{fig:acceptances}
\end{figure}

The simulated samples of \four and \two events are used to evaluate the total signal efficiencies \Afourmu and \Atwomu, respectively, 
given by the product of detector geometric acceptance and reconstruction and selection efficiencies, as functions of \pt for two 
rapidity regions: $\abs{y} < 1.5$ and $1.5 < \abs{y} < 2.4$. Figure~\ref{fig:acceptances} shows the total efficiencies \vs \pt, split 
by rapidity range. In two-muon decays, \Atwomu is limited by the trigger efficiency, reaching a plateau of about 70\%. In contrast, in 
four-muon decays, the efficiency to reconstruct all four muons in the event has a maximum value of about 15\%. The low-\pt behavior is 
understood to be correlated to the minimum \pt of about 3.5\GeV required for a muon in the central region to reach the muon detectors. 
The high-\ptfourmu drop off, in turn, comes from the difficulty of reconstructing four muons with very small angular separation, 
owing to the boost of the parent \e meson.

The branching fraction $\Bfourmu \equiv \mathcal{B}(\four)$ is determined relative to $\Btwomu \equiv \mathcal{B}(\two)$ using 
\begin{linenomath}
\begin{equation}
    \frac{\Bfourmu}{\Btwomu} = \frac{\Nfourmu}{ \sum\limits_{\mathrm{i,j}} \Ntwomu \frac{\Afourmu}{\Atwomu}},
    \label{eq:BR}
\end{equation}
\end{linenomath}
\noindent where \Nfourmu is the total four-muon signal yield, \Ntwomu are the two-muon signal yields in the bins $\mathrm{i}$ and $\mathrm{j}$ 
of the candidate \e meson's \pt and rapidity, and \Afourmu and \Atwomu are the corresponding efficiencies. We define 32 bins in \pt in 
the range 7--70\GeV and 2 bins in $\abs{y}$. Because of the relative nature of the measurement, several uncertainties cancel out when 
considering the ratio of quantities. Remaining uncertainties are assessed for each component of Eq.~(\ref{eq:BR}). The uncertainty 
on \Btwomu is 14\%~\cite{PDG2022}, while the statistical uncertainty on \Nfourmu is estimated to be 16\% from the fit shown in 
Fig.~\ref{fig:fourMuonMassFitData}. A similar source of uncertainty would arise from the measurement of \Ntwomu, but it is negligible 
compared to the other uncertainties when considering the large sample of \two decays. 

The remaining dominant uncertainties are systematic in nature and arise from incomplete knowledge of the efficiencies evaluated by 
simulation. This type of uncertainty is subdivided into three parts: (i) on the track \pt threshold, 9.0\%; (ii) on the trigger \pt 
threshold, 8.4\%; and (iii) on the efficiency plateau, 3.2\%. Parts (i) and (ii) are caused by imperfect modeling of the turn-on behavior 
of the single-muon reconstruction efficiency observed in data. They are estimated by varying the thresholds in simulation and measuring 
the corresponding variation of the relative \Nfourmu yield. A conservative but meaningful range of threshold variations is determined 
for (i) by comparing the momentum scale variation that is required to shift the mass peak of the \e meson by the width observed in the 
two-muon spectrum. For (ii), we assume a 10\% uncertainty in the effective threshold of the trigger. The uncertainty on (iii) is determined 
by measuring the trigger efficiency in data with an unbiased sample of events collected with electron triggers. The discrepancy between 
this efficiency in data and simulation leads to a 3.2\% correction to the branching fraction, which is also taken as the corresponding 
uncertainty. This uncertainty affects both decay channels in an unbalanced fashion owing to the different number of final-state muons, 
and hence it only partially cancels out in the ratio. A subdominant source of systematic uncertainty is attributed to the choice of fit 
model used to extract the signal yield in both channels. This uncertainty is assessed by testing several alternative signal and background 
models, and determining the variation in signal yield, resulting in a value of 6.6\%. Overall, we estimate the total systematic uncertainty 
in the branching ratio measurement to be 14\%, adding all contributions in quadrature.

The resulting $\Bfourmu/\Btwomu$ ratio is:
\begin{linenomath}\begin{equation*}
\frac{\Bfourmu}{\Btwomu} = [ 0.86 \pm 0.14 \stat \pm 0.12 \syst ] \times 10^{-3}.
\end{equation*}\end{linenomath}
With $\mathcal{B}(\two) = (5.8 \pm 0.8) \times 10^{-6}$~\cite{PDG2022}, the branching fraction of the newly observed four-muon decay channel is measured to be
\begin{linenomath}\begin{equation*}
	\mathcal{B}(\four) = [ 5.0 \pm 0.8 \stat \pm 0.7 \syst \ \pm 0.7 \experr ] \times 10^{-9},
\end{equation*}\end{linenomath}
where the last term reflects the uncertainty in $\mathcal{B}(\two)$. The measurement is in agreement with the theoretical 
prediction of $(3.98 \pm 0.15) \times 10^{-9}$~\cite{escribanoDatadrivenApproachPi2018}.

In summary, the first observation of the \e meson's rare double-Dalitz decay to four muons is reported. This is made possible by the use 
of CMS data collected with high-rate muon triggers at $\sqrt{s} = 13\TeV$ in 2017 and 2018, corresponding to an integrated luminosity 
of 101\fbinv. The branching fraction of the \four decay is measured relative to the \two decay, yielding a ratio of branching fractions 
of $[ 0.86 \pm 0.14 \stat \pm 0.12 \syst ] \times 10^{-3}$. Using the world average branching fraction value~\cite{PDG2022} for the normalization 
channel, the branching fraction of the four-muon decay channel is 
$\mathcal{B}(\emmmm) = [ 5.0 \pm 0.8 \stat \pm 0.7 \syst \pm 0.7 \experr ] \times 10^{-9}$, where the last term is 
the uncertainty in the normalization channel branching fraction. This result is in agreement with theoretical predictions~\cite{escribanoDatadrivenApproachPi2018}. The augmented statistical power of data collected with high-rate triggers in future runs of the LHC will enable a detailed study of doubly virtual pseudoscalar transition form factors (TFFs) in the time-like region~\cite{Escribano_2015}. Measuring TFFs is crucial to understand the effects of the strong interaction in a range of relevant processes, such as the anomalous magnetic moment of the muon~\cite{Gan:2020aco}.

\begin{acknowledgments}
We congratulate our colleagues in the CERN accelerator departments for the excellent performance of the LHC and thank the technical and administrative staffs at CERN and at other CMS institutes for their contributions to the success of the CMS effort. In addition, we gratefully acknowledge the computing centres and personnel of the Worldwide LHC Computing Grid and other centres for delivering so effectively the computing infrastructure essential to our analyses. Finally, we acknowledge the enduring support for the construction and operation of the LHC, the CMS detector, and the supporting computing infrastructure provided by the following funding agencies: BMBWF and FWF (Austria); FNRS and FWO (Belgium); CNPq, CAPES, FAPERJ, FAPERGS, and FAPESP (Brazil); MES and BNSF (Bulgaria); CERN; CAS, MoST, and NSFC (China); MINCIENCIAS (Colombia); MSES and CSF (Croatia); RIF (Cyprus); SENESCYT (Ecuador); MoER, ERC PUT and ERDF (Estonia); Academy of Finland, MEC, and HIP (Finland); CEA and CNRS/IN2P3 (France); BMBF, DFG, and HGF (Germany); GSRI (Greece); NKFIH (Hungary); DAE and DST (India); IPM (Iran); SFI (Ireland); INFN (Italy); MSIP and NRF (Republic of Korea); MES (Latvia); LAS (Lithuania); MOE and UM (Malaysia); BUAP, CINVESTAV, CONACYT, LNS, SEP, and UASLP-FAI (Mexico); MOS (Montenegro); MBIE (New Zealand); PAEC (Pakistan); MES and NSC (Poland); FCT (Portugal); ENMAK (Turkey); NASU (Ukraine); STFC (United Kingdom); DOE and NSF (USA).
\end{acknowledgments}

\bibliography{auto_generated} 
\cleardoublepage \appendix\section{The CMS Collaboration \label{app:collab}}\begin{sloppypar}\hyphenpenalty=5000\widowpenalty=500\clubpenalty=5000
\cmsinstitute{Yerevan Physics Institute, Yerevan, Armenia}
{\tolerance=6000
A.~Hayrapetyan, A.~Tumasyan\cmsAuthorMark{1}\cmsorcid{0009-0000-0684-6742}
\par}
\cmsinstitute{Institut f\"{u}r Hochenergiephysik, Vienna, Austria}
{\tolerance=6000
W.~Adam\cmsorcid{0000-0001-9099-4341}, J.W.~Andrejkovic, T.~Bergauer\cmsorcid{0000-0002-5786-0293}, S.~Chatterjee\cmsorcid{0000-0003-2660-0349}, K.~Damanakis\cmsorcid{0000-0001-5389-2872}, M.~Dragicevic\cmsorcid{0000-0003-1967-6783}, A.~Escalante~Del~Valle\cmsorcid{0000-0002-9702-6359}, P.S.~Hussain\cmsorcid{0000-0002-4825-5278}, M.~Jeitler\cmsAuthorMark{2}\cmsorcid{0000-0002-5141-9560}, N.~Krammer\cmsorcid{0000-0002-0548-0985}, D.~Liko\cmsorcid{0000-0002-3380-473X}, I.~Mikulec\cmsorcid{0000-0003-0385-2746}, J.~Schieck\cmsAuthorMark{2}\cmsorcid{0000-0002-1058-8093}, R.~Sch\"{o}fbeck\cmsorcid{0000-0002-2332-8784}, D.~Schwarz\cmsorcid{0000-0002-3821-7331}, M.~Sonawane\cmsorcid{0000-0003-0510-7010}, S.~Templ\cmsorcid{0000-0003-3137-5692}, W.~Waltenberger\cmsorcid{0000-0002-6215-7228}, C.-E.~Wulz\cmsAuthorMark{2}\cmsorcid{0000-0001-9226-5812}
% ...
\par}
\vskip\cmsinstskip
\dag:~Deceased\\
$^{1}$Also at Yerevan State University, Yerevan, Armenia\\
$^{2}$Also at TU Wien, Vienna, Austria\\
% ...
\end{sloppypar}
\end{document}